\documentclass[12pt]{article}

\usepackage[T1]{fontenc}

\usepackage{datetime}

\usepackage[letterpaper, nohead, left=1.0in, right=1.0in, top=1.0in, bottom=1.0in, dvips ]{geometry}

\usepackage{cite}

\usepackage[labelfont=bf]{caption}
\usepackage[lofdepth,lotdepth]{subfig}
\usepackage{graphicx,color}

\begin{document}

\begin{center}
{\large Investigating the Use of Mastery-Style Online Homework Exercises \par in Introductory Algebra-based Mechanics in a Controlled Clinical Study }\par 
{William R. Evans and Mats A. Selen}\par
{Department of Physics, University of Illinois at Urbana-Champaign}\par
\end{center}

\begin{abstract}
Homework in introductory physics represents an important part of a student's learning experience; therefore choosing the manner in which homework is presented merits investigation.  We performed three rounds of clinical trials comparing the effects of mastery-style homework vs traditional-style homework with students in both algebra-based and calculus-based introductory mechanics. Results indicate a benefit from mastery-style over traditional-style homework, principally for weaker students who are less familiar with the material being covered and on questions that are nearer transfer to the study materials.  
\end{abstract}

\section{Introduction}
Homework represents a significant part of a student's learning resources in many introductory physics classes, so optimizing the pedagogy of online homework delivery is an important goal of education research.  While there are many options for homework delivery, the one that we are investigating is the mastery method of learning, pioneered by Benjamin Bloom in the 1960s \cite{Bloom1968}. 

The mastery method is a cycle of formative assessment followed by additional learning opportunities.  When a student completes a series of homework problems, correctness feedback is given on the set of problems as a whole. If the student reaches a certain correctness threshold (often 100\%), the student is considered to have mastered the material and is permitted to move on to the next section or topic.  If the student does not master the material, they are given additional learning opportunities to help them improve their understanding.  These additional learning opportunities are followed by a similar set of homework problems on the same material. This cycle continues until the student has achieved mastery. 

Mastery-inspired homework has been tried in several different course settings.  One such setting was in an introductory electricity and magnetism course for physics and engineering majors at the University of Illinois.  Gladding, et al \cite{NoahFall2014Phys212Study} compared computer-based mastery-style homework with a more traditional computer-based homework format currently used in the introductory physics classes at the University of Illinois.  This traditional format involves immediate correctness feedback and an unlimited number of attempts.  In the Gladding, et al study, students who completed homework presented in the mastery style significantly outperformed students who completed the same homework presented in the more traditional style on a post-test covering the topics that the students studied in the homework.  

Motivated by the positive effects of mastery-style homework observed in the study by Gladding, et al, the primary goal of this study was to investigate whether this pedagogy would have a similar effect on students in an algebra-based introductory physics class, which primarily serves life science majors.  The question of how best to teach physics to life science students has been a topic of conversation for decades, and has resulted in various types of reformed Introductory Physics for Life Science (IPLS) courses at different institutions (see References \cite{Redish2009, Redish-Poster-2012, Redish-CBE-2013, Redish2014, O'Shea2013} for some representative reforms).  While most studies of reformed IPLS courses have focused on the content being taught in the course, this study focused on the pedagogy by which that content is delivered.

\section{Overview of Clinical Trials}

The goal of this experiment was to evaluate the effectiveness of computer-based mastery-inspired homework for learning in introductory mechanics in a controlled clinical setting.  Three clinical trials were run in this experiment.  A summary of the students participating in each trial is given in Table \ref{ParticipantTable}.  In each trial, participating students were randomly divided into two groups.  One group received a series of homework problems in the traditional format described above.  The other group received the same series of homework problems in the mastery format. 

\begin{table}[h]
   \centering
   \begin{tabular}{| p{1.0cm} | p{1.7cm} | p{1.6cm} | p{1.5cm} | p{2.5cm} | p{2.6cm} | p{2.6cm} |}
      \hline
      \bf{Trial} & \bf{Course} & \bf{Topic} & \bf{Timing} & \bf{Total} \newline \bf{Participants} & \bf{Traditional- Homework Participants} & \bf{Mastery- Homework Participants} \\ \hline \hline
      1 & Algebra- \newline based \newline Mechanics & Friction & Prior to \newline First Midterm & 31 & 15 & 16 \\ \hline
      2 & Calculus- \newline based \newline Mechanics & Friction & Prior to \newline First Midterm & 31 & 15 & 16 \\ \hline
      3 & Algebra- \newline based \newline Mechanics & Newton's \newline Laws of \newline Motion & Prior to \newline Final Exam & 49 & 19 & 30 \\ \hline
   \end{tabular}
   \caption{Summary of participants in each of the three clinical trials.}
   \label{ParticipantTable}
\end{table}

Students in this experiment were enrolled either in the algebra-based \emph{College Physics: Mechanics and Heat} or in the calculus-based \emph{University Physics: Mechanics} courses at the University of Illinois at Urbana-Champaign.  The algebra-based course is required for many of the life science majors on campus, and most of the students in this course are juniors and seniors.  The calculus-based course is the first in a sequence of introductory physics courses required for physics and engineering majors at the university.  It is generally taken during the freshman year of a student's college career. 

No course credit was offered to students for participating in these experimental trials. Rather, these trials were advertised as a means of helping students prepare for an upcoming exam.  Participating students were told that they would receive extra office hour help from experienced TAs after the experiment. 

In the mastery-style homework condition, students were presented with sets of four or five questions on a given topic. Once all the questions in a set were completed, the student submitted these questions to be graded, and correctness feedback was given. At this point, these students had access to narrated, animated solution videos to the problems they had just completed.  These solution videos constituted the additional learning opportunities outlined in the description of the mastery method of learning.  These videos were designed to walk students through the concepts used in the problems.  The goal of these videos was to help the students complete subsequent problem sets on the same topic.  If a student completed all of the questions in the problem set correctly, they were said to have mastered the topic of that problem set and were permitted to move on to the next topic.  If a student did not complete all the questions in the problem set correctly, they were then presented with an equivalent problem set on the same topic. The problems in these subsequent sets were designed to be similar, but not identical, to the original problems. This cycle continued until the student completed all the questions in a problem set correctly. Because students in the mastery-style homework condition had to complete an entirely new set if any question was answered incorrectly, mastery-style questions were given in a multiple-choice format to help students catch minor errors and minimize the resulting student frustration.  

In the traditional-style homework condition, students had access to all the topics simultaneously, so questions could be completed in any order. Students received correctness feedback on their answers after each submission.  An unlimited number of submissions were allowed for each question. The problem sets viewed by the students in the traditional-style homework condition were the initial problem sets on each topic viewed by the students in the mastery-style homework condition.  Since students in this group had immediate correctness feedback and an unlimited number of tries, the traditional-style homework problems were presented as free-response questions rather than multiple-choice.  To balance the additional help that the students in the mastery-style homework condition received from the solution videos, students doing homework problems in the traditional style were encouraged to use course notes and to work with each other in completing their problem sets. These resources are similar to those a student might use when doing their regular homework in the course. 

After students in both groups completed the homework problems, they were given a paper-based free-response post-test on the same topics covered in the homework problems. Each of the twelve questions on the post-test was graded out of three points by two different graders.  In cases where the scores given by the two graders differed by more than 1 point, the student's response was re-visited and discussed by both graders until a consensus could be reached.  The average of the scores given by the two graders was used in the subsequent analysis.

\section{First Clinical Trial} 

\begin{figure}[h]
      \centering
         \includegraphics[width=0.7\textwidth]{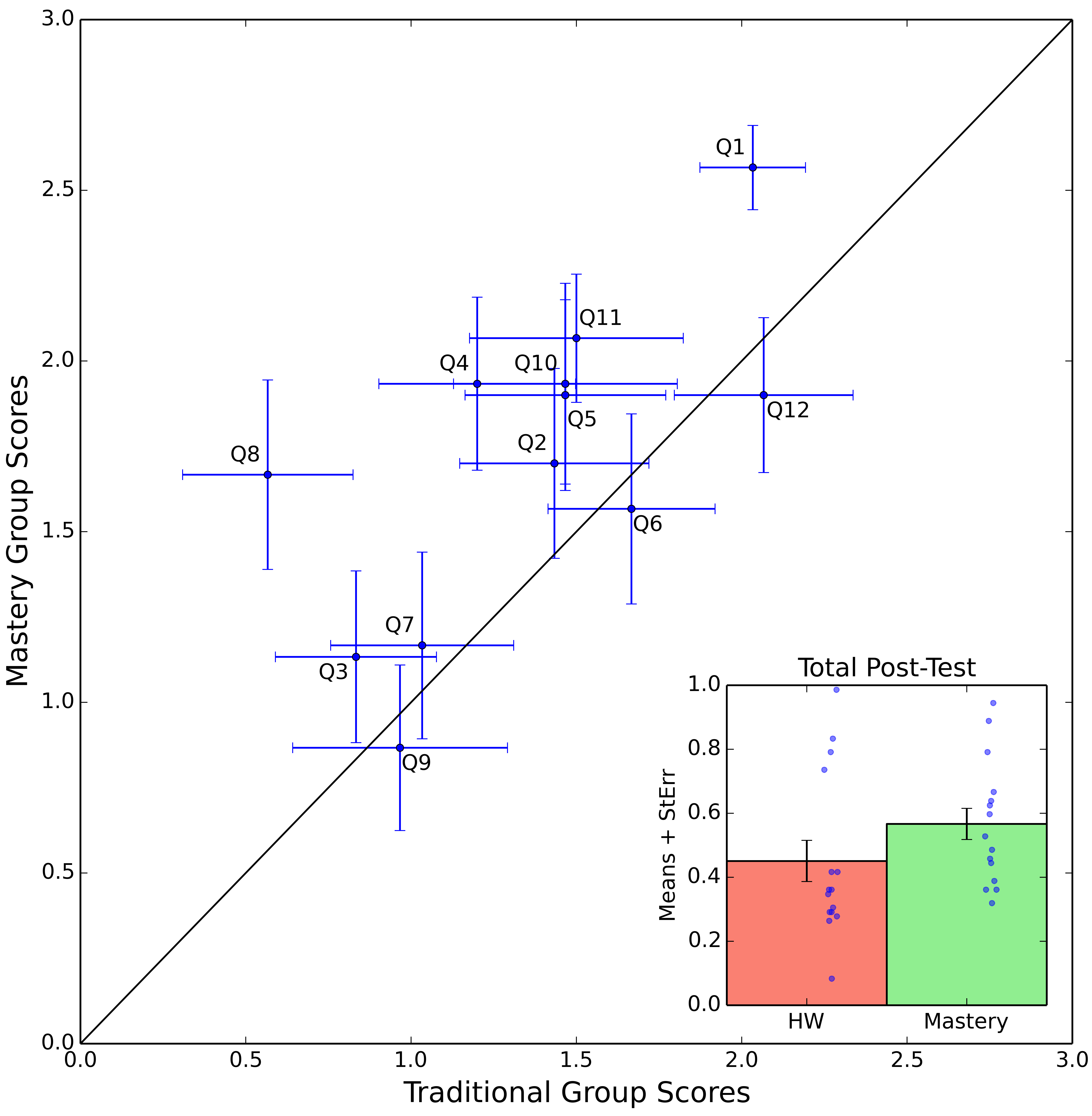}
         \captionsetup{width=1.0\textwidth}
         \caption{Normalized average scores on each of the twelve questions on the post-test used in the first clinical trial.  Average scores with standard errors for students in the mastery-homework condition are plotted on the vertical axis, and average scores with standard errors for students in the traditional-homework condition are plotted on the horizontal axis.  Points appearing above the diagonal line indicate questions on which the students in the mastery-homework condition outperformed students in the traditional-homework condition.\\  \underline{Inset:}  Normalized total average scores on the post-test with standard errors are shown for the two homework conditions.  Dots on the inset graph represent individual students' total scores.  Compare Figure 4 from Reference \cite{NoahFall2014Phys212Study}.}
         \label{Clinical1_101OverallFig}
   \end{figure}

The first trial involved students from algebra-based physics and was run shortly before the students' first midterm exam (see Table \ref{ParticipantTable}).  Figure \ref{Clinical1_101OverallFig} shows the scores from each group on each of the twelve questions on the post-test.  

On average, students in the mastery homework condition outperformed students in the traditional homework condition on 9 of the 12 questions on the post-test, significantly so on 3 of them ($p<0.05$, one-tail \emph{t}-test).  There were no questions on the post-test on which the students in the traditional homework condition significantly outperformed the students in the mastery homework condition.  The post-test questions on which the students in the mastery-homework condition most significantly outperformed the students in the traditional-homework condition appeared to be those that were most similar to the homework questions.  

\begin{figure}[h]
      \centering
         \subfloat[Subfigure 1 list of figures text][Performance of students on those post-test questions which were judged to be nearer-transfer to the homework problems.  The difference between the mastery and traditional homework groups was judged to be significant with \emph{p} < 0.05.]{
            \includegraphics[width=0.45\textwidth]{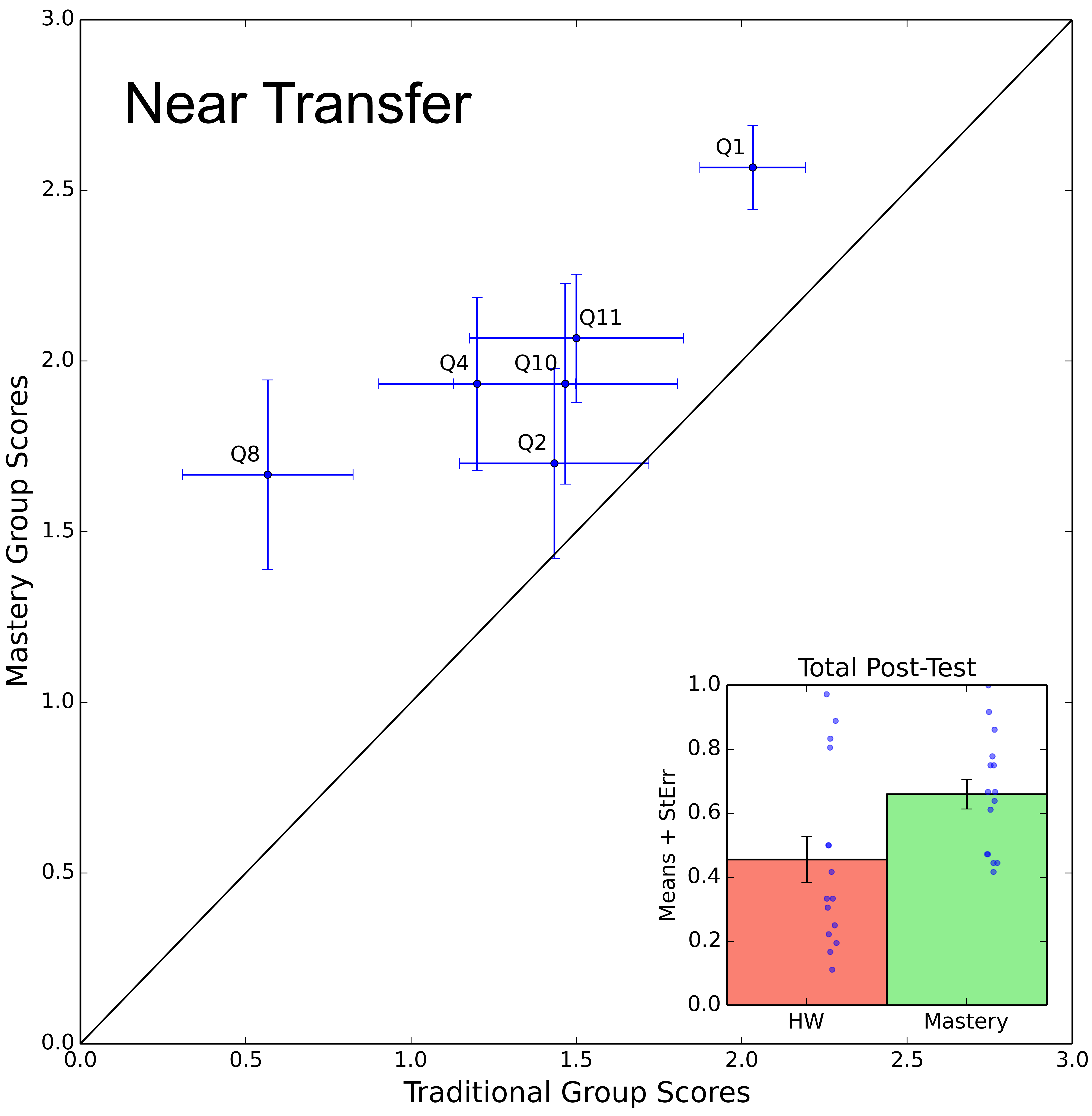}
            \label{Clinical1_101NearFig}
         }
         \qquad
         \subfloat[Subfigure 2 list of figures text][Performance of students on those post-test questions which were judged to be farther-transfer to the homework problems.  The difference between the mastery and traditional homework groups was not significant.]{
            \includegraphics[width=0.45\textwidth]{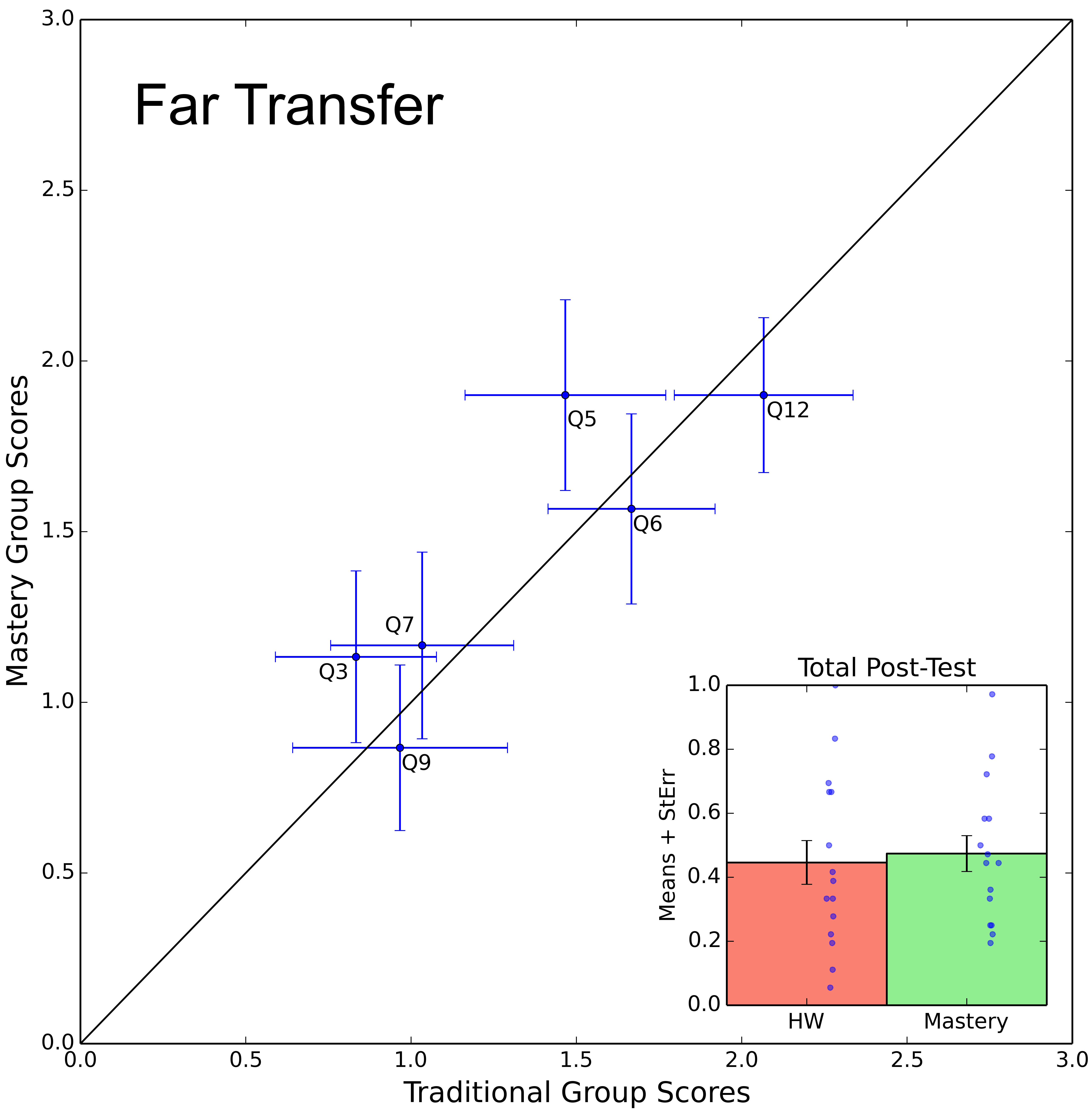}
            \label{Clinical1_101FarFig}
         }
         \caption{Performance of students in the first clinical trial, with nearer- and farther-transfer questions analyzed separately.  }  
         \label{Clinical1_101NearAndFarFigs}
   \end{figure}

To test the hypothesis that a problem's nearness of transfer was correlated with the two groups' difference in performance on that problem, we presented the post-test and the homework from the clinical study to eight physics education experts at the University of Illinois.  These experts rated the post-test questions on a scale of 0 to 5 by how near or far transfer the questions were from the homework problems.  Their responses were averaged, and the questions were split into two equal-sized groups representing those problems with nearer or farther degrees of transfer.  The results are shown in Figure \ref{Clinical1_101NearAndFarFigs}.  

Figure \ref{Clinical1_101NearFig} shows that the students in the mastery-homework condition significantly outperformed the students in the traditional-homework condition on the nearer-transfer questions from the post-test.  This difference in performance was significant on 3 of the 6 questions ($p<0.05$, one-tail t-test), and marginally significant on another one ($p<0.1$, one-tail t-test).  The difference in performance between students in the two homework conditions was not judged to be significant on any of the six farther-transfer questions (see Figure \ref{Clinical1_101FarFig}).  

   \begin{figure}[h]
      \centering
         \subfloat[Subfigure 1 list of figures text][Performance of students in the bottom half of each homework condition group as measured by their final semester course scores.  The difference in post-test total scores is significant with $p<0.05$.]{
            \includegraphics[width=0.45\textwidth]{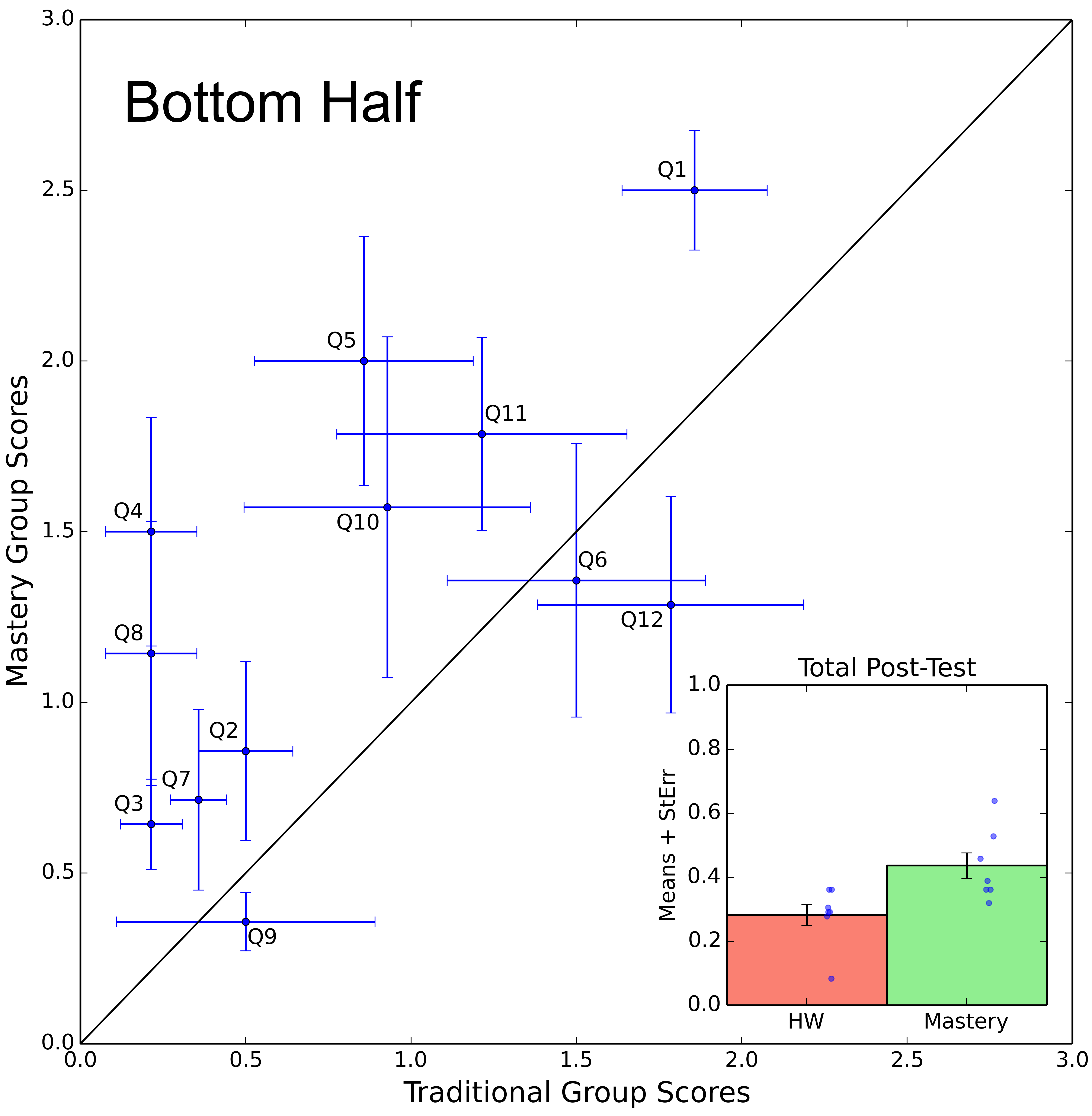}
            \label{Clinical1_101BottomHalfFig}
         }
         \qquad
         \subfloat[Subfigure 2 list of figures text][Performance of students in the top half of each homework condition group as measured by their final semester course scores.  The difference in post-test total scores is not significant.]{
            \includegraphics[width=0.45\textwidth]{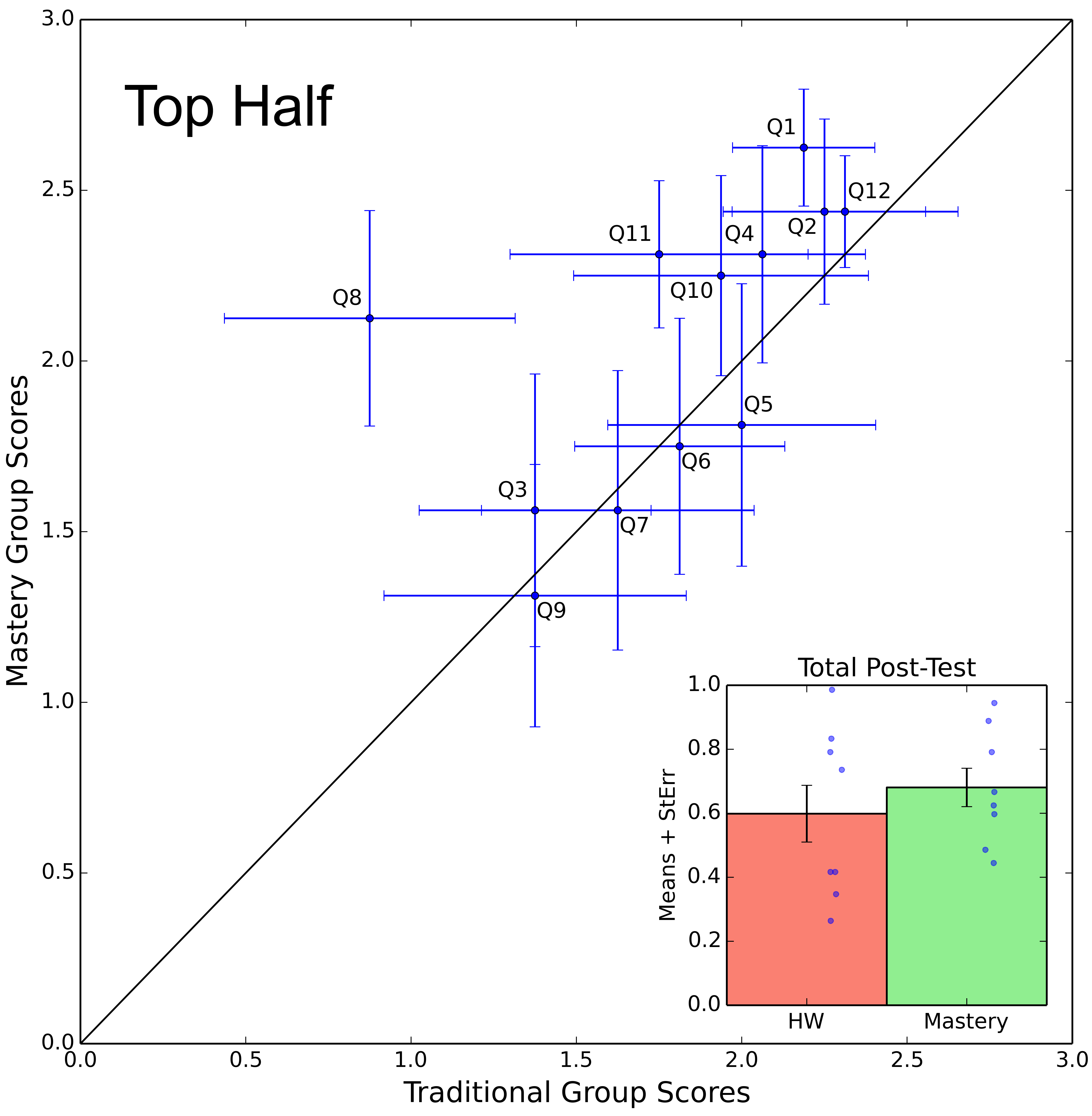}
            \label{Clinical1_101TopHalfFig}
         }
         \caption{Performance of students in the first clinical trial with students in the top and bottom half of each homework condition group analyzed separately.  This median split was performed using the students' final semester course scores.  Dots on the inset graphs show individual student post-test totals.}  
         \label{Clinical1_101BottomAndTopHalfFigs}
   \end{figure}

In addition, we analyzed the effects of mastery-style homework on the stronger and weaker students in each group separately.  Figure \ref{Clinical1_101BottomAndTopHalfFigs} shows the performance of students in the top and bottom half of each homework condition group, as measured by their final semester course scores.  As can be seen in Figure \ref{Clinical1_101BottomHalfFig}, the difference in post-test performance between the mastery-homework group and the traditional-homework group is highly significant for the weaker students.  Weaker students in the mastery-homework condition significantly outperformed the weaker students in the traditional-homework condition on the post-test ($p<0.05$).  For comparison, Figure \ref{Clinical1_101TopHalfFig} shows that the stronger students in each homework condition performed almost identically on the post-test.

\section{Second and Third Clinical Trials}

The second trial involved students from calculus-based physics and was run shortly before the first midterm exam (see Table \ref{ParticipantTable}).  We used the same materials that were used in the first trial with the students in algebra-based mechanics, which allowed us to directly compare the performance of the students in the two classes.  Students in the calculus-based class demonstrated higher proficiency with problems involving friction, in that their average scores on the post-test were higher (74\% $\pm$ 3\%) than the students in the algebra-based course (50\% $\pm$ 4\%).  

\begin{figure}[h]
      \centering
         \subfloat[Subfigure 1 list of figures text][Performance of students in the second clinical trial.  This trial involved students in calculus-based physics and covered the topic of friction, using the same materials as the first clinical trial.  This trial was run before the first midterm in the course.  The difference in total scores is not significant.]{
            \includegraphics[width=0.45\textwidth]{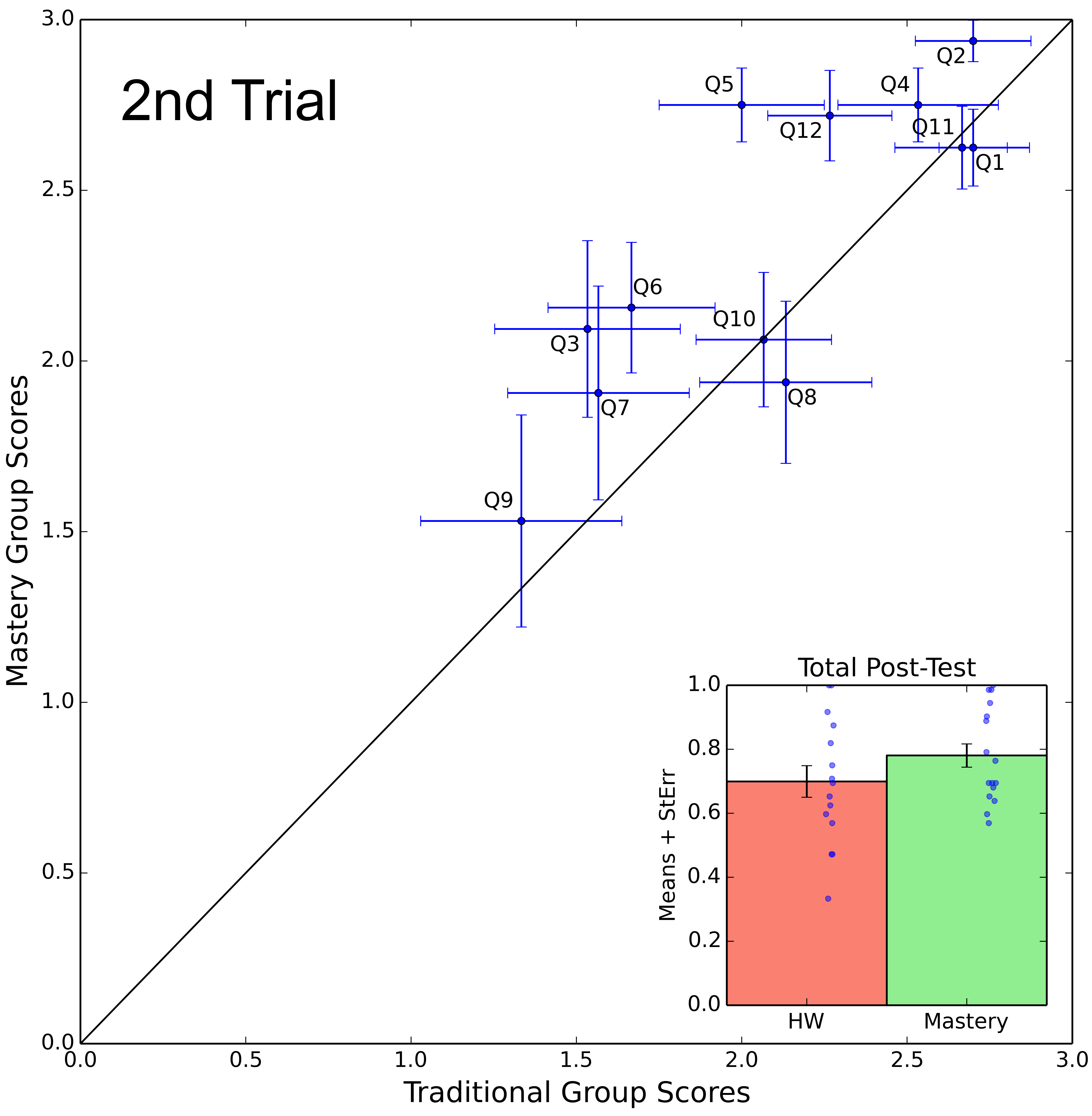}
            \label{Clinical2_211OverallFig}
         }
         \qquad
         \subfloat[Subfigure 2 list of figures text][Performance of students in the third clinical trial.  This trial involved students in algebra-based physics and covered the topic of Newton's Laws.  The clinical trial was run before the course's comprehensive final exam.  The difference in total scores is not significant.]{
            \includegraphics[width=0.45\textwidth]{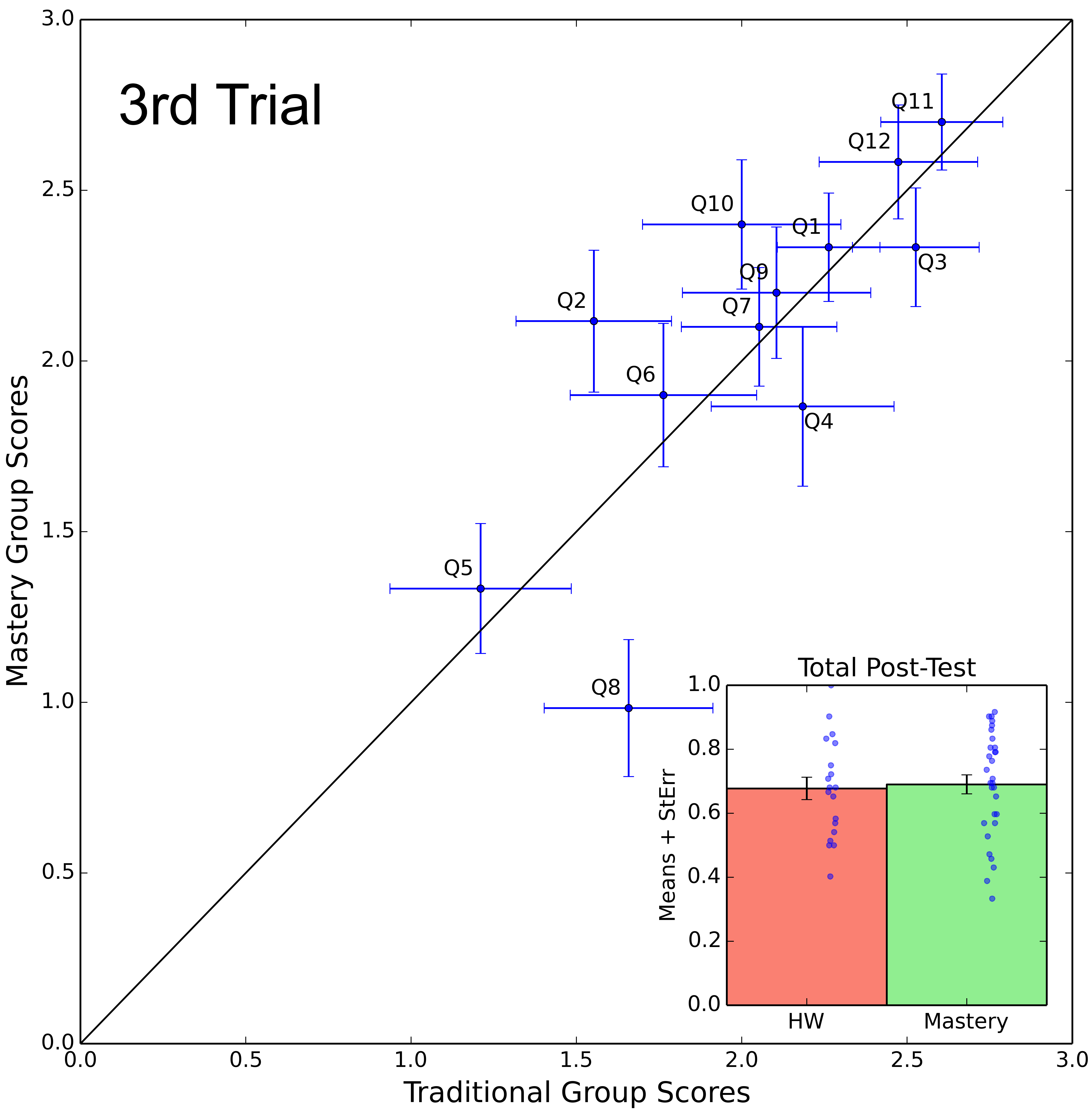}
            \label{Clinical3_101OverallFig}
         }
         \caption{Performance of students in the second and third clinical trials.  Average scores with standard errors for students in the mastery-homework condition are plotted on the vertical axis, and average scores with standard errors for students in the traditional-homework condition are plotted on the horizontal axis.  Average total scores for the students in each homework condition are shown in the inset.  Dots on the inset graphs show individual student post-test totals.}  
         \label{Clinicals2-3_OverallFigs}
   \end{figure}

In this trial, no difference was seen when comparing students in the mastery-homework condition with the students in the traditional-homework condition.  Student performance on the post-test in the second trial is shown in Figure \ref{Clinical2_211OverallFig}.  

The third trial involved students from algebra-based physics and covered the subject of Newton's laws of motion (see Table \ref{ParticipantTable}). This trial was run shortly before the final exam, by which point the students had been using Newton's laws for most of the semester. No difference was seen when comparing students in the mastery-homework condition with the students in the traditional-homework condition in this trial.  Student performance on the post-test in the third trial is shown in Figure \ref{Clinical3_101OverallFig}.  

\section{Discussion}

The results of the first clinical trial show that the mastery-style homework approach can have a positive impact on students in algebra-based physics.  Although our statistics were limited, significant differences between the treatment and control groups were seen on near-transfer problems and for students least proficient with the material. 

The results of the second clinical trial, in which engineering and physics majors were given the same treatment and control conditions used with the life science majors in the first trial, showed no difference between the two homework conditions.  The results of the third clinical trial, which was given at the end of the semester and which focused on Newton's laws of motion, also showed no difference between the treatment and control groups.  Both of these results are consistent with the hypothesis that students already proficient with the material benefit less from mastery-style homework than students that are initially less proficient with the material.  The uncertainties, however, are too large to make a significant claim. 

While further work is needed to verify the effects of mastery-style homework on learning in introductory physics, our results show that mastery-style homework appears to benefit some students in introductory mechanics over traditional-style homework.  This effect is most significant with students who are weaker or less familiar with the material, or when the degree of transfer between the learning materials and the testing materials is small.  While there is likely no ``silver bullet'' to help students overcome all the difficulties in learning physics, these results indicate that appropriately used mastery-style homework activities can help students better master the skills needed in algebra-based introductory mechanics.

\begin{center}
   \textbf{Acknowledgment}
\end{center}

This work was partially funded by a grant from the Strategic Instructional Innovations Program at the University of Illinois.  The authors would also like to thank the University of Illinois Physics Education Research Group, especially Brianne Gutmann and Elaine Schulte for particularly helpful conversations and editing.  

\bibliographystyle{ieeetr}
\bibliography{PERBibliography}{}

\begin{thebibliography}{1}

\bibitem{Bloom1968}
B.~S. Bloom, ``Learning for mastery,'' {\em Evaluation Comment}, vol.~1,
  pp.~1--12, May 1968.

\bibitem{NoahFall2014Phys212Study}
G.~Gladding, B.~Gutmann, N.~Schroeder, and T.~Stelzer, ``Clinical study of
  student learning using mastery style versus immediate feedback online
  activities,'' {\em Physical Review Special Topics -- Physics Education
  Research}, vol.~11, no.~010114, 2015.

\bibitem{Redish2009}
E.~F. Redish and D.~Hammer, ``Reinventing college physics for biologists:
  Explicating and epistemological curriculum,'' {\em American Journal of
  Physics}, vol.~77, 2009.

\bibitem{Redish-Poster-2012}
E.~F. Redish, B.~W. Dreyfus, B.~D. Geller, V.~Sawtelle, J.~Svoboda, and
  C.~Turpen, ``Developing a research-based interdisciplinary physics course for
  biologists,'' 2012.
\newblock Contributed poster and talk, AAPT National Meeting, Philadelphia, PA,
  July 28-Aug 3, 2012.

\bibitem{Redish-CBE-2013}
E.~F. Redish and T.~J. Cooke, ``Learning each other's ropes: Negotiating
  interdisciplinary authenticity,'' {\em CBE -- Life Sciences Education},
  vol.~12, pp.~175--186, June 2013.

\bibitem{Redish2014}
E.~F. Redish, C.~Bauer, K.~L. Carleton, T.~J. Cooke, M.~Cooper, C.~H. Crouch,
  B.~W. Breyfus, B.~Geller, J.~Giannini, J.~S. Gouvea, M.~W. Klymkowsky,
  W.~Losert, K.~Moore, J.~Presson, V.~Sawtelle, K.~V. Thompson, C.~Turpen, and
  R.~K.~P. Zia, ``{NEXUS}/{P}hysics: An interdisciplinary repurposing of
  physics for biologists,'' {\em American Journal of Physics}, vol.~82, 2014.

\bibitem{O'Shea2013}
B.~O'Shea, L.~Terry, and W.~Benenson, ``From ${F}=ma$ to flying squirrels:
  Curricular change in an introductory physics course,'' {\em CBE -- Life
  Sciences Education}, vol.~12, pp.~230--238, June 2013.

\end{thebibliography}

\end{document}